\documentclass{aa}  
\usepackage[utf8]{inputenc}
\usepackage{graphicx}
\usepackage{natbib}
\usepackage{adjustbox}
\usepackage{txfonts}
\usepackage{array}
\usepackage{threeparttable}
\usepackage{lineno}
\usepackage{newtxtext,newtxmath}
\usepackage{multirow}
\newcommand{\HI}{H\,{\sc i}}

\usepackage[autostyle]{csquotes}
\usepackage{textcomp}
\usepackage{rotating}
\usepackage{float}
\MakeRobust{\overrightarrow}
\usepackage{tikz}
\usetikzlibrary{tikzmark, calc}
\usepackage{hyperref}
\usepackage{enumitem}
\usepackage{subcaption}
\usepackage{amssymb,amsmath}
\usepackage{pifont}
\usepackage[T1]{fontenc}

\hypersetup{
    colorlinks=true,
    linkcolor=blue,
    filecolor=magenta,      
    urlcolor=blue,
    citecolor=blue,
    pdftitle={Overleaf Example},
    pdfpagemode=FullScreen,
    }
\usepackage{calc}  

\usepackage{booktabs}  
\begin{document} 

\titlerunning{Sextans~B with MeerKAT}
\authorrunning{Namumba et al.}
   \title{Ram-pressure signatures in the dwarf irregular galaxy Sextans~B revealed by deep MeerKAT \HI\ observations}

\author{Brenda Namumba\inst{1}\thanks{bnamumba@iaa.es}
          \and 
           Neel Kolhe\inst{2}
          \and
          Francois Hammer\inst{2}
          \and
          Yanbin Yang\inst{2}    
          \and
          Claude Carignan\inst{3,4,5}
          \and
          Roger Ianjamasimanana\inst{1}
          \and
          Gyula I. G. Józsa\inst{6,7}
          \and
          Lourdes Verdes-Montenegro\inst{1}
          \and
           Haifeng Wang\inst{8}
          \and
          Hao Chen\inst{9}
          \and
          Xin Huang\inst{2}
          \and
          Fortune Ndalama\inst{10}
          \and
          Amidou Sorgho\inst{1}
          \and
          Marie Korsaga\inst{1,5}
          \and
           Saul P. Phiri \inst{10}
}
\institute{
        Instituto de Astrofísica de Andalucía (CSIC), Glorieta de la Astronomía s/n, 18008 Granada, Spain
       \and
        LIRA, Observatoire de Paris, Université PSL, CNRS, 92190 Meudon, France
        \and
       Department of Astronomy, University of Cape Town, Private Bag X3, Rondebosch 7701,South Africa
       \and
       Département de physique, Université de Montréal, Complexe des sciences MIL, 1375 Avenue
       Thérèse-Lavoie-Roux, Montréal, QC, Canada H2V 0B3
       \and
       Observatoire d’Astrophysique de l’Université Ouaga I Pr Joseph Ki-Zerbo (ODAUO), BP 7021, Ouaga 03, Burkina Faso
       \and
       Max-Planck-Institut für Radioastronomie Radioobservatorium Effelsberg, Max-Planck-Strasse 28 53902 Bad Münstereife,  Germany
       \and
       Centre for Radio Astronomy Techniques and Technologies (RATT), Department of Physics and Electronics, Rhodes University, Makhanda 6140, South Africa
       \and
       Local Universe and Time-Domain Astronomy Laboratory, Department of Astronomy, China West Normal University, Nanchong, 637002, China
       \and
       Research Center for Computational Earth and Space Science, Zhejiang Laboratory, Hangzhou 311121, China
       \and
       The Copperbelt University, Jambo Drive, Riverside, P.o.Box 21692, Kitwe, ZAMBIA 
        }
   \date{Received 03 04, 2026; accepted 16 06, 2026}

\abstract{
\textbf{Context.} The impact of extremely low-density environments such as the diffuse intergalactic medium (IGM) on the neutral gas distribution of dwarf galaxies remains poorly explored observationally. 

\textbf{Aims.} We present deep MeerKAT \HI\ 21 cm observations of the Local Group dwarf irregular galaxy Sextans~B that achieve a spectral resolution of $1.4\,\mathrm{km\,s^{-1}}$ and reach column-density sensitivities down to $N_{\mathrm{HI}} \sim 3.3 \times 10^{18}\,\mathrm{cm^{-2}}$, allowing us to trace the extended \HI\ disc and faint outer structures with high sensitivity.

\textbf{Methods.} We analysed the \HI\ distribution and compared it with the stellar component. Three-dimensional kinematic modelling of the \HI\ cube was performed using TiRiFiC. We performed hydrodynamical simulations tailored to Sextans~B, which show that the IGM ram-pressure acting on the outer gas disc can produce asymmetric gas distributions, filamentary structures, and kinematic perturbations.

\textbf{Results.} The low-column-density \HI\ distribution is asymmetric and reveals a remarkable filamentary structure arranged in a rosette superposed on the \HI\ disc. A comparison with the stellar distribution shows spatial offsets between the gaseous and stellar components, with the stellar disc remaining relatively symmetric and the \HI\ envelope becoming increasingly disturbed. Three-dimensional kinematic modelling of the \HI\ cube using TiRiFiC reproduces the global velocity gradient but reveals systematic differences between the approaching and receding sides of the rotation curve at large radii, indicating that the outer velocity field departs from simple axisymmetric rotation. While stellar feedback can produce small-scale cavities and turbulence in dwarf galaxies, it cannot generate the \HI\ filamentary structure, the large-scale asymmetric outer \HI\ envelope, or the systematic divergence between the approaching and receding rotation curves observed here. This is consistent with the effects expected from interaction with a diffuse IGM. 

\textbf{Conclusions.} The combination of morphological and kinematic signatures therefore suggests that the outer \HI\ disc of Sextans~B is affected by a ram-pressure interaction with the diffuse IGM in the outskirts of the Local Group. This is the second strong example in the Local Group, after WLM, showing that even a very low-density IGM can significantly influence the gas distribution and kinematics in the outer parts of dwarf galaxies.
     } 


   \maketitle
\nolinenumbers
\section{Introduction}
Dwarf galaxies in the Local Group experience a wide range of environmental conditions, ranging from satellites embedded within the halos of massive galaxies to systems residing in the low-density outskirts of the group \citep{2012AJ....144....4M}. Owing to their shallow gravitational potentials, dwarf irregular galaxies are particularly sensitive to external perturbations, making them valuable probes of environmental processes operating across a wide range of gas-density regimes. Environmental effects on dwarf galaxies have long been studied in dense environments such as galaxy clusters, where ram-pressure stripping and tidal interactions can strongly affect the structure and dynamics of the interstellar medium \citep{2006PASP..118..517B,2020MNRAS.494.1114S}. Similar environmental processes have also been investigated within the circumgalactic halos of massive galaxies, including the Milky Way, where interactions with the surrounding medium may influence the evolution of dwarf systems \citep{Wang2024}. In contrast, the potential influence of extremely low-density environments, such as the diffuse intergalactic medium (IGM) that surrounds the Local Group, remains much less explored observationally.

Environmental interactions occur within several distinct gaseous media characterised by markedly different densities. These environments form a hierarchy that includes the dense intracluster medium (ICM) in galaxy clusters, the intra-group gas medium in galaxy groups, the circumgalactic medium (CGM) surrounding massive galaxies,  and finally the extremely diffuse intergalactic medium (IGM). In rich galaxy clusters, interaction with the intracluster medium (ICM; $n \gtrsim 10^{-3}$--$10^{-2}\,\mathrm{cm^{-3}}$) can efficiently remove large fractions of a galaxy’s interstellar medium through classical ram-pressure stripping \citep{1972ApJ...176....1G,2006PASP..118..517B}. Such processes are thought to play an important role in the transformation of dwarf galaxies in clusters. 

In galaxy groups and around massive galaxies such as the Milky Way, satellite systems are embedded within extended hot circumgalactic halos (CGM; $n \sim 10^{-5}$-- 2 $10^{-4}\,\mathrm{cm^{-3}}$ at galactocentric distances of tens to hundreds of kiloparsecs \citep{Kalberla2006,Grcevich2009,2013ApJ...770..118M}. This halo environment is thought to play a central role in shaping the evolution of satellite galaxy populations. This influence is reflected in the strong contrast between the gas-poor dwarf spheroidal satellites surrounding the Milky Way and the gas-rich dwarf irregular galaxies typically found at larger distances from the Galaxy, consistent with the well-known morphology--density relation \citep{2012AJ....144....4M}. The CGM is also thought to influence large gaseous structures such as the Magellanic Stream \citep{2015ApJ...813..110H,Wang2019}.

These environments span several orders of magnitude in gas density and therefore correspond to different regimes of hydrodynamical interaction, from strong stripping in clusters to weaker perturbations in group and intergalactic environments. Between the CGM of individual galaxies and the extremely diffuse intergalactic medium, dwarf galaxies located in loose groups may also interact with a low-density intra-group medium with typical densities of $n \sim 10^{-6}$--$10^{-5}\,\mathrm{cm^{-3}}$. Beyond the virial radii of massive galaxies, the gas density declines further into the intergalactic medium (IGM) of the Local Group, where typical densities are expected to be significantly lower ($n \lesssim 10^{-6}$--$10^{-5}\,\mathrm{cm^{-3}}$; \citep{2002AJ....123.1316B,2022A&A...660L..11Y}).  In this low-density regime, the dynamical impact on isolated dwarf galaxies remains uncertain, although they have been explored by \citet{Kolhe2026}.

Internal processes such as stellar feedback are widely recognised as the primary drivers of small-scale structure in the interstellar medium of dwarf irregular galaxies, producing cavities, shells, and turbulent gas motions in the neutral gas \citep{1992AJ....103.1841P,1999AJ....118..273W,2011AJ....141...23B,2015ApJ...813..110H}. However, because dwarf galaxies possess shallow gravitational potentials, their outer low–column-density gas is particularly vulnerable to external perturbations. 

Deep interferometric studies of nearby dwarf galaxies have revealed a variety of \HI\ structures, including asymmetric envelopes, one-sided truncations, gas--star offsets, and filamentary extensions that may potentially reflect environmental interactions \citep{2002AJ....123.1316B,2021ApJ...913...53P}. Although many of these features are typically attributed to internal processes such as stellar feedback or turbulence \citep{2021ApJ...913...53P}, interactions with a diffuse ambient medium may also contribute to shaping the outer gas morphology in some systems \citep{2002AJ....123.1316B,2022A&A...660L..11Y}. However, direct observational evidence for a significant influence of such low-density media on the \HI\ structure of isolated dwarf galaxies remains limited.

The possibility that the extremely low-density intergalactic medium may affect both the gas distribution and the kinematics of isolated dwarf irregular galaxies has only recently begun to be explored. Deep MeerKAT observations of the isolated dwarf galaxy WLM revealed extended low-column-density \HI\ features aligned with its motion, interpreted as consistent with mild ram-pressure interaction with a diffuse Local Group medium \citep{2022A&A...660L..11Y}. Subsequent kinematic analysis revealed systematic differences between the approaching and receding sides of the disc, suggesting that interaction with a diffuse medium may also perturb the outer velocity field \citep{Kolhe2026}. These results suggest that even very weak ram pressure, acting over long timescales, may gradually influence the outer gas discs of dwarf galaxies in low-density environments.

Galaxies located in the outskirts of the Local Group therefore provide suitable laboratories for testing this scenario, as tidal interactions are expected to be weak while sensitivity to diffuse ambient gas remains high. Sextans~B is a gas-rich dwarf irregular galaxy located at a distance of 1.36 Mpc \citep{2005AJ....130.1558K}. It lies in the Local Group outskirts near other dwarf galaxies such as Sextans~A and NGC~3109, which are located at similar distances of approximately $1.3$--$1.4$ Mpc \citep{2003AJ....125.1261D,2006ApJ...648..375S}, but do not constitute a bound galaxy group. Its relative isolation minimizes the likelihood of strong tidal perturbations, placing it in a regime where any environmental influence would most plausibly result from interaction with a low-density circumgalactic, intra-group, or intergalactic medium.

Previous \HI\ studies of Sextans~B reported an extended rotating gas disc with a largely regular global morphology and a rising rotation curve \citep{2012AJ....144..134H,2018MNRAS.478..487N}. KAT-7 observations identified global kinematic differences between the approaching and receding sides of the disc, together with multiple \HI\ holes attributed to stellar feedback \citep{2018MNRAS.478..487N}. However, the limited angular resolution and surface-brightness sensitivity of earlier data prevented a detailed investigation of the faint, low-column-density outer gas, where signatures of weak environmental interaction are expected to emerge.

The deep MeerKAT observations presented here reach \HI\ column densities below $10^{18}\,\mathrm{cm^{-2}}$ and provide sub-kiloparsec resolution, enabling us to trace the extended \HI\ disc of Sextans~B, identify faint asymmetric and filamentary structures, compare the distributions of stars and neutral gas, and perform three-dimensional kinematic modelling of the data cube. By combining morphological and dynamical signatures with hydrodynamical expectations for weak ram-pressure interaction, we test whether the observed properties of Sextans~B are consistent with interaction between its \HI\ disc and a diffuse ambient medium in the outskirts of the Local Group.

The structure of the paper is as follows: Section~\ref{sec:sample} describes relevant properties of the target galaxy, followed by Section~\ref{sec:observations}, which details observations and data reduction. Section~\ref{sec:distribution} presents the \HI\ morphology, comparison with stellar distributions, and filament analysis. Section~\ref{sec:kinematics} discusses the three-dimensional kinematic modelling. Section~\ref{sec:simulations} compares the observations with ram-pressure simulations, and Section~\ref{sec:discussion} discusses the implications for dwarf galaxy evolution. Section~\ref{sec:conclusion} summarises the conclusions.

\section{Target galaxy: Sextans~B}\label{sec:sample}
Sextans~B (DDO~70, UGC~5373) was selected as the target of this study to investigate the impact of low-density environmental processes on the neutral gas distribution and kinematics of dwarf irregular galaxies. Sextans~B is a gas-rich dwarf irregular galaxy of morphological type IBm \citep{1991rc3..book.....D} located on the outskirts of the Local Group, near the galaxies NGC~3109 and Sextans~A. At a distance of $1.36 \pm 0.07$~Mpc \citep{2005AJ....130.1558K}, it resides in a regime where strong tidal interactions with massive galaxies are expected to be minimal, making it well suited for investigating subtle environmental effects such as interaction with a diffuse intra-group or intergalactic medium.

Previous \HI\ studies revealed an extended neutral gas disc with a largely regular morphology and a rising rotation curve, together with multiple \HI\ holes attributed to stellar feedback \citep{2012AJ....144..134H}. However, the limited angular resolution and surface-brightness sensitivity of previous observations prevented a detailed investigation of faint, low-column-density gas in the outer regions of the disc, where signatures of weak environmental interaction are expected to be detectable. 

Sextans~B is therefore a suitable target for probing possible low-level ram-pressure effects in a low-density environment, as its relative isolation minimises tidal perturbations while its extended \HI\ disc provides enhanced sensitivity to interaction with surrounding diffuse gas. 
Deep optical imaging from the Large Binocular Telescope (LBT) in the $g$ and $r$ bands \citep{2014A&A...566A..44B}, where details of the observational setup and exposure times are provided, provides resolved stellar density maps, enabling a direct comparison between the spatial distributions of stars and neutral hydrogen. These data, combined with the high sensitivity and sub-kiloparsec spatial resolution of MeerKAT, allow us to trace faint outer \HI\ structures, identify possible filamentary features, and perform detailed three-dimensional kinematic modelling of the \HI\ data cube, which was not possible with previous observations. The adopted global properties of Sextans~B used in this work are summarised in Table~\ref{tsextans_parameters}.

\begin{table}[ht]
\centering
\caption{Key parameters of Sextans~B. Each parameter is labeled and the corresponding references are provided in the notes.}
\label{tsextans_parameters}
\begin{tabular}{lcc}
\hline
\hline
Parameter & Sextans~B & Label \\
\hline
RA (J2000) & 10:00:00.10 & (a) \\  
DEC (J2000) & $+05$:19:55.99 & (a) \\  
Distance (Mpc) & $1.36 \pm 0.07$ & (b) \\
Stellar Mass ($M_\odot$) & $(4-5) \times 10^7$ & (b) \\
\HI\ Mass ($M_\odot$) & $4.8 \times 10^7$ & (c) \\
\HI\ diameter (3$\sigma$) (arcmin) & 20 & (c) \\
Systemic Velocity (km s$^{-1}$) & 301 & (c) \\
Morphology & IBm & (d) \\
\hline
\end{tabular}
\begin{tablenotes}
\small
\item Notes: References corresponding to each label: 
(a) \citep{2017ApJS..233...25A}; 
(b) \citep{2005AJ....130.1558K}; 
(c) \citep{2018MNRAS.478..487N}; 
(d) \citep{1991rc3..book.....D}.
\end{tablenotes}
\end{table}
\section{MeerKAT observations and data reduction}\label{sec:observations}
\HI\ 21-cm line observations of Sextans~B were obtained with the MeerKAT radio telescope \citep{2016mks..confE...1J,2018ApJ...856..180C} under project SCI-20220822-BN-01 (PI: B. Namumba). The galaxy was observed using the L-band receivers for approximately three hours of on-source integration time on 15--16 December 2022. Observations were carried out using the NE107M narrowband mode, which provides a 107~MHz bandwidth with 3.3~kHz channel spacing ($\approx0.7$~km\,s$^{-1}$). Before data reduction, data were spectrally averaged over two channels, resulting in a final spectral resolution of 1.4~km\,s$^{-1}$.

Data reduction was performed using the CARAcal pipeline \citep{2020ascl.soft06014J}, which includes standard processing steps such as radio-frequency interference (RFI) flagging, calibration, continuum subtraction, and imaging. Automated RFI flagging was carried out with AOFlagger \citep{2010ascl.soft10017O,2012A&A...539A..95O}. Bandpass and flux calibration were derived from the primary calibrator using the MeerKAT flux-density scale, whereas time-dependent complex gains were obtained from interleaved phase-calibrator scans. Data quality was verified through visual inspection of diagnostic plots at each processing stage. Continuum emission was removed in the $uv$-plane using a first-order polynomial fit to line-free channels.

Imaging of the \HI\ data cubes was performed with WSClean \citep{2014MNRAS.444..606O,2017MNRAS.471..301O} within CARAcal. A two-step imaging strategy was adopted. First, a low-resolution cube with a $90''$ Gaussian $uv$ taper was produced to generate a source mask using SoFiA-2 \citep{2021MNRAS.506.3962W}. This mask was subsequently applied during deconvolution of the final data cubes, where multiscale cleaning was used to ensure sensitivity to emission on all spatial scales. To fully exploit MeerKAT's angular resolution and surface-brightness sensitivity, three cubes were produced with beam sizes ranging from $\sim9''$ to $67''$, using Briggs robust weighting ($r = 0$ and $r = 0.5$) combined with Gaussian $uv$ tapering. The resulting datasets reach \HI\ column-density sensitivities of $N_{\mathrm{HI}} \sim 10^{20}$--$10^{18}$~cm$^{-2}$ ($3\sigma$, over 20~km\,s$^{-1}$). The observational setup and imaging parameters are summarised in Table~\ref{observations}.

Moment maps for Sextans~B were derived from the calibrated \HI\ data cubes using SoFiA-2. Signal detection employed the smooth+clip algorithm, with spatial kernels at the native resolution and 4 pixels, and spectral kernels at the native resolution, 9 channels ($\sim12.6$~km\,s$^{-1}$), and 25 channels ($\sim35$~km\,s$^{-1}$). A clipping threshold of $4\sigma$ was applied, with detections required to span at least five spatial pixels and eight spectral channels. The $4\sigma$ threshold was adopted to minimise contamination from noise peaks while preserving low-column-density emission in the outer disc. A reliability threshold of 0.8 was imposed to retain only statistically significant emission. Zeroth-, first-, and second-moment maps were generated from the resulting masks. A stricter intensity mask was applied when computing the moment-1 and moment-2 maps to minimise spurious high-velocity and high-dispersion pixels introduced by noise in low signal-to-noise regions. The moment maps are presented in Section~\ref{sec:distribution}.
\begin{table*}[htbp]
\centering
\caption{Summary of observations and properties of \HI\ cubes for Sextans~B.}
\label{observations}
\begin{tabular}{lc}
\toprule
Number of antennas & 58 to 64 \\
Total integration (target) & 3 h on source \\
FWHM of primary beam & $\sim$1$^\circ$ \\
Calibrated channel width & 6.4 kHz (1.4 km/s) \\
\bottomrule
\end{tabular}

\vspace{1em} 

\begin{tabular}{l l r r r r}
\toprule
Galaxy & Cube & Resolution ($^{\prime\prime} \times ^{\prime\prime}$) & Pixel size ($^{\prime\prime}$) & 1$\sigma_\mathrm{rms}$ (mJy/beam) & $N_\mathrm{HI}$ (3$\sigma$, 20 km/s) [cm$^{-2}$] \\
\midrule
Sextans~B & t=0, r=0    & 10.36 $\times$ 7.53  & 3  & 0.72 & 1.63$\times10^{20}$ \\
           & t=0, r=0.5  & 15.52 $\times$ 10.19 & 3  & 0.58 & 6.49$\times10^{19}$ \\
           & t=30, r=0.5 & 46.32 $\times$ 41.03 & 5  & 0.72 & 6.68$\times10^{18}$ \\
           & t=60, r=0.5 & 69.38 $\times$ 65.36 & 10 & 0.86 & 3.34$\times10^{18}$ \\
\bottomrule
\end{tabular}

\smallskip
\noindent{\footnotesize
\textit{Notes.}
1 $\sigma_\mathrm{rms}$ measured at the velocity resolution of 1.4 km\,s$^{-1}$. \\
$r$ is the Briggs robust weighting parameter, and $t$ is the Gaussian taper applied to the $uv$ data. \\
t=0, r=0.5 corresponds to the high-resolution cube.
}
\end{table*}
\section{\HI\ distribution and morphology}\label{sec:distribution}
\begin{figure} 
\centering
   \advance\leftskip0cm
     \includegraphics[width=8.5cm]{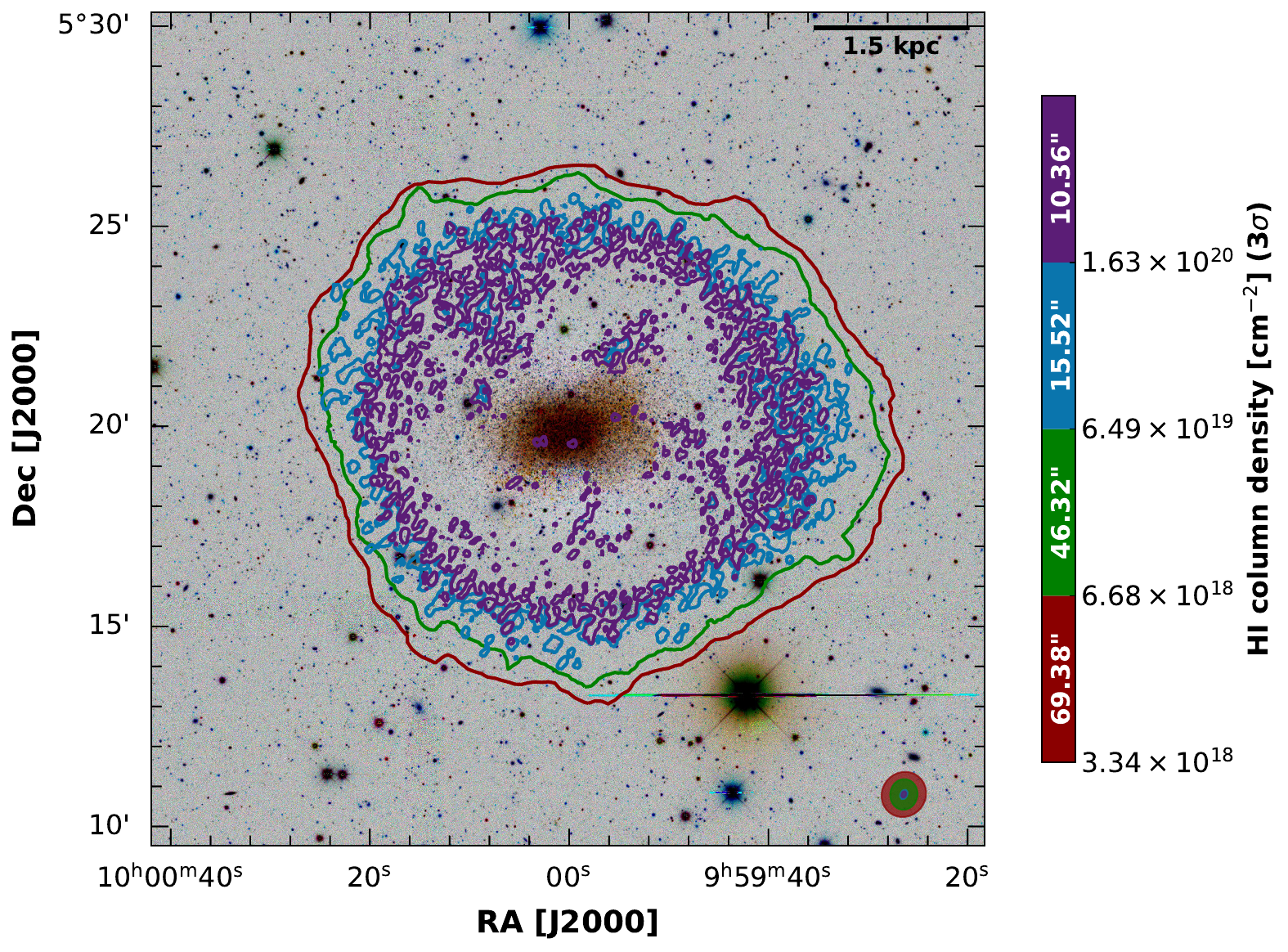}
\caption{MeerKAT \HI\ column density contours from multiple angular-resolution cubes of Sextans~B overlaid on DECaLS gri composite images. This optical image is included for visual reference only and is not a deep optical map (see Fig.~\ref{fig:high} for the deeper imaging used for comparison). Each contour corresponds to the $3\sigma$ column density sensitivity of the respective cube. The colour bar indicates the beam size associated with each resolution. The ellipses in the lower-right corner show the synthesised beams, while the black horizontal line indicates the physical scale in kpc.}
\label{fig:mom0}
\end{figure}

The MeerKAT observations reveal an extended \HI\ disc in Sextans~B that is detected over a wide range of spatial scales (Fig.~\ref{fig:mom0}). To search for low-column-density features in the outer regions, we primarily use the low-resolution \HI\ column-density maps, which are optimised for surface-brightness sensitivity. The lowest-resolution cube ($\sim69''$ beam) reaches a $3\sigma$ sensitivity of $N_{\mathrm{HI}} \approx 3.3\times10^{18}$~cm$^{-2}$ over 20~km\,s$^{-1}$, enabling the detection of diffuse neutral gas at large galactocentric radii. At these column densities, the \HI\ disc extends well beyond the main stellar body, reaching projected radii of $\sim1.5$~kpc along the major axis. This corresponds to approximately twice the optical half-light radius of Sextans~B ($r_e \approx 0.75$~kpc; \citealt{2014A&A...566A..44B}), adopting the same distance as in this work. This increased extent at lower resolution reflects the use of a larger beam that smooths the emission and reduces small-scale structure, making the same gas distribution appear more extended, while the total flux remains unchanged. At different spatial resolutions, the \HI\ distribution shows a systematic asymmetry across the disc, with the emission extending further towards the west than on the opposite side. 
\subsection{MeerKAT high-resolution \HI\ maps}
\begin{figure*} 
\centering
   \advance\leftskip0cm
     \includegraphics[width=15.5cm]{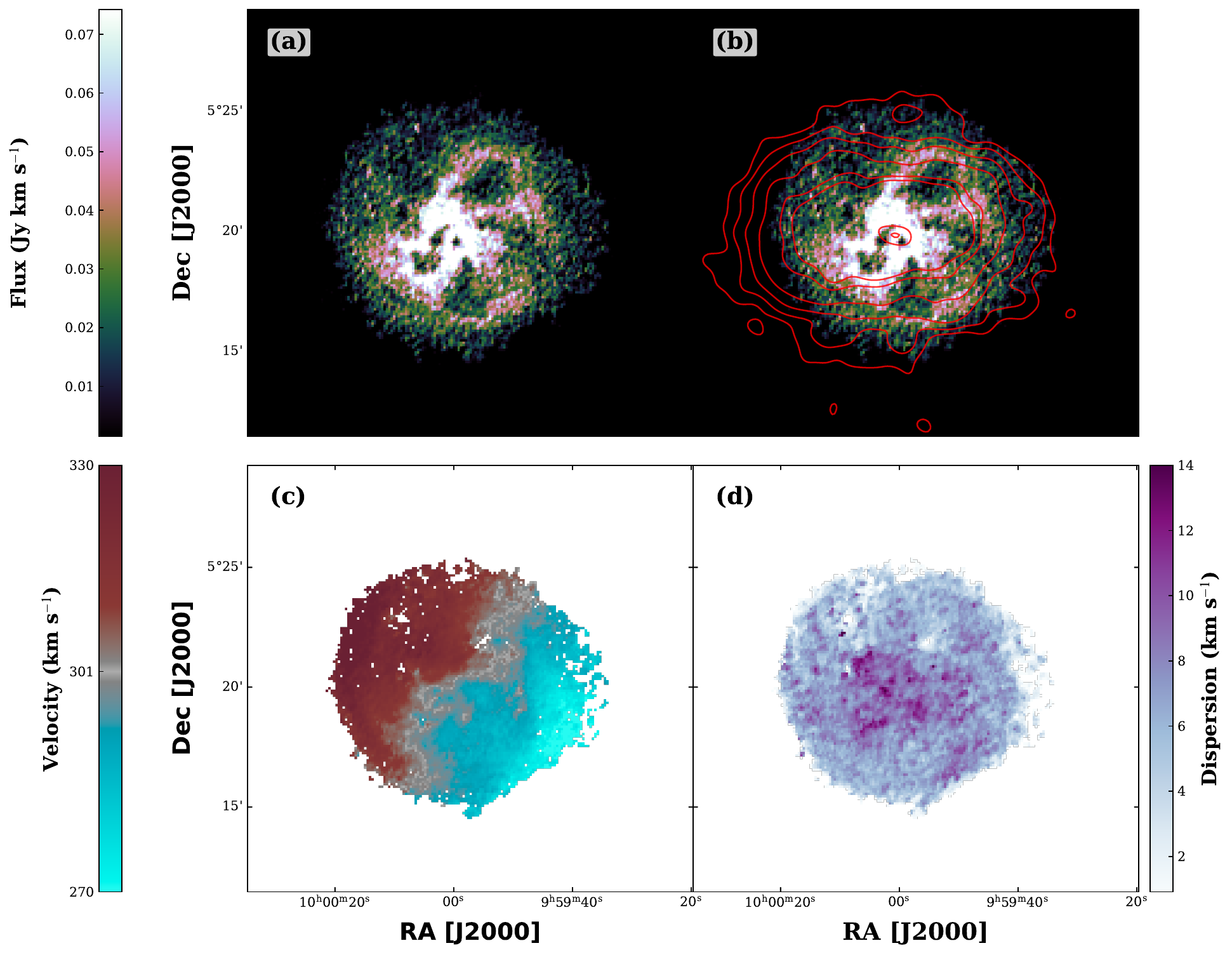}
\caption{Highest-resolution MeerKAT \HI\ maps of Sextans~B. Panel (a): integrated \HI\ intensity (moment~0). Panel (b): optical stellar emission (red contours, in arbitrary units) overlaid on the \HI\ moment~0 map; the contours are at 5, 10, 20, 40, 80, and 100$\sigma$ above the background noise ($\sigma$) of the optical image. Panel (c): intensity-weighted velocity field (moment~1). Panel (d): velocity dispersion map (moment~2). The maps show a centrally concentrated \HI\ disc with pronounced internal structure, including multiple cavities embedded in clumpy high-column-density gas, together with small-scale structure extending into the outer regions.}
\label{fig:high}
\end{figure*}

The highest-resolution MeerKAT maps (Fig.~\ref{fig:high}) reveal the detailed internal structure of the \HI\ disc in Sextans~B. The integrated intensity map (panel~a) shows a centrally concentrated distribution with pronounced substructure, including several well-defined \HI\ cavities embedded within clumpy high-column-density gas. These cavities are irregular in shape and size, with typical diameters of approximately $0.7$--$0.75$~kpc. The gas forms a prominent central peak surrounded by patchy emission and structured density gradients. Such substructures are likely related to stellar feedback processes, although external perturbations or gas inflow cannot be excluded as contributing factors. This fine-scale morphology, characterised by significant internal complexity and localised clumps, was not resolved in the previous lower-resolution KAT-7 observations \citep{2018MNRAS.478..487N}.

\begin{figure} 
\centering
   \advance\leftskip0cm
   \includegraphics[width=7.2cm]{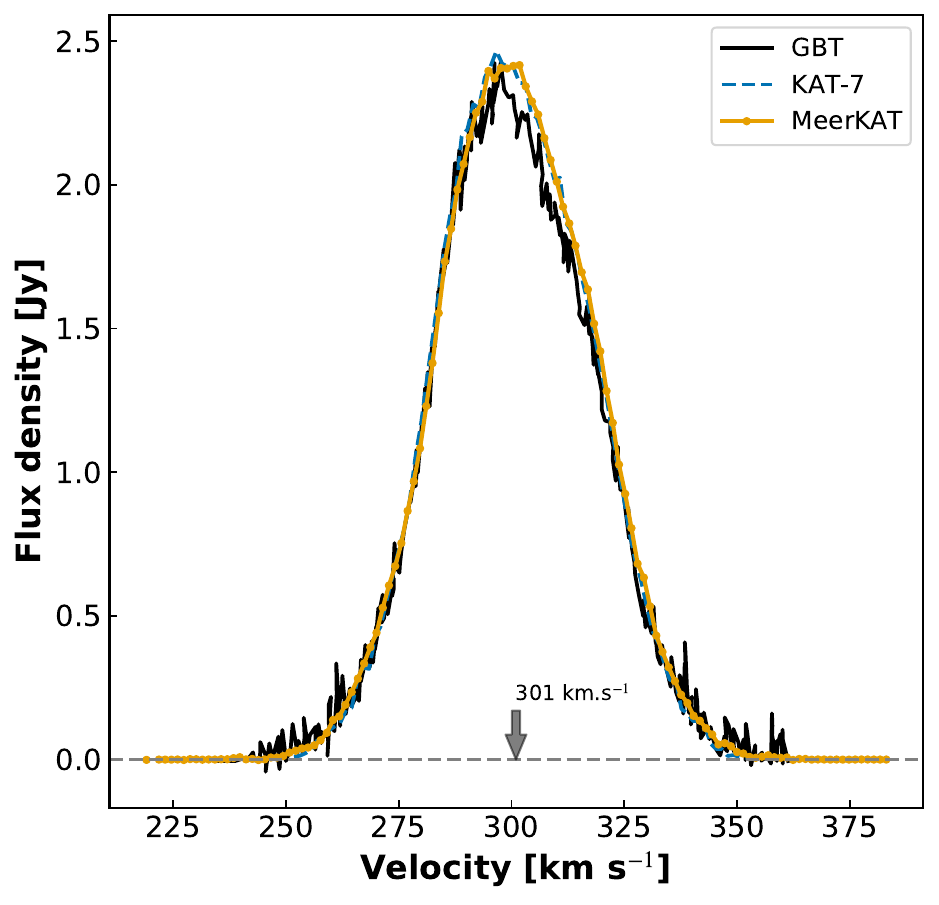}
\caption{Integrated \HI\ spectrum of Sextans~B from MeerKAT (orange), compared with KAT-7 \citep[blue;][]{2018MNRAS.478..487N} and single-dish measurements \citep[black;][]{2005ApJS..160..149S,2012AJ....144..134H}. The arrow indicates the systemic velocity derived in Section~\ref{sec:kinematics}. The horizontal dashed grey line marks the zero level. The profile is extracted from the mid-resolution data cube.}
\label{fig:profile}
\end{figure}

At this resolution, \HI\ emission is detected across most of the disc, with the outer regions appearing increasingly fragmented. The intensity-weighted velocity field (panel~c) exhibits a smooth large-scale velocity gradient across the galaxy, with localised distortions in the inner disc that spatially coincide with regions of complex \HI\ structure, including cavities, clumpy emission, and locally enhanced velocity dispersion. The velocity dispersion map (panel d) shows values ranging from $\sim6$ to $\sim30$~km\,s$^{-1}$. A detailed analysis of the \HI\  kinematics based on the moment~1 and moment~2 maps is presented in Section~\ref{sec:kinematics}.
The integrated \HI\ spectrum of Sextans~B is shown in Fig.~\ref{fig:profile}. The profile is relatively narrow and symmetric. The MeerKAT data yield an integrated flux of $104.5 \pm 1.5$~Jy~km~s$^{-1}$, corresponding to an \HI\ mass of $(4.6 \pm 0.07)\times10^{7}\,M_\odot$ at a distance of 1.36~Mpc.
The \HI\ mass was computed using the relation from \citet{1997AJ....114.1858P}. 
\[
M_{\mathrm{HI}} = 2.36 \times 10^{5} \, D^{2} \, \int S_v \, dv,
\]
where $D$ is the distance in Mpc and $\int S_v \, dv$ is the integrated flux density in Jy\,km\,s$^{-1}$. We measure linewidths of $W_{20}=59.5 \pm 1.4$~km~s$^{-1}$ and $W_{50}=38.8 \pm 2.1$~km~s$^{-1}$, which are consistent with previous KAT-7 measurements \citep{2018MNRAS.478..487N}. A comparison of the global \HI\ properties derived from KAT-7 and MeerKAT is presented in Table~\ref{tab:kat7_meerkat_comparison}.

\begin{table}[ht]
\centering
\small
\setlength{\tabcolsep}{5pt}
\renewcommand{\arraystretch}{1.0}
\begin{threeparttable}
\caption{Comparison of \HI\ properties of Sextans~B from KAT-7 and MeerKAT}
\label{tab:kat7_meerkat_comparison}
\begin{tabular}{lcc}
\toprule
Parameter & KAT-7 & MeerKAT \\
\midrule
Integrated flux (Jy\,km\,s$^{-1}$) & $105.0 \pm 1.4$ & $104.5 \pm 1.5$ \\
Distance (Mpc)                     & 1.36 & 1.36 \\
\HI\ mass ($10^{7}\,M_\odot$) & $4.60 \pm 0.06$ & $4.60 \pm 0.07$ \\
$W_{50}$ (km\,s$^{-1}$)            & $40.8 \pm 1.8$ & $38.8 \pm 2.1$ \\
$W_{20}$ (km\,s$^{-1}$)            & --- & $59.5 \pm 1.4$ \\
\bottomrule
\end{tabular}
\begin{tablenotes}
\item[] \footnotesize Notes: The KAT-7 measurements are from \citet{2018MNRAS.478..487N}. \HI\ masses are calculated using our adopted distance of 1.36 Mpc.
\end{tablenotes}
\end{threeparttable}
\end{table}
\subsection{Comparison between the \HI\ and stellar discs}\label{sec:histars}
Fig.~\ref{fig:high} Panel (b) shows optical stellar density contours (red) adapted from \cite{Bellazzini2014}, overlaid on the highest-resolution \HI\ integrated intensity map. Although the stellar and \HI\ components share a common central peak, their outer distributions differ markedly. The stellar disc appears relatively smooth and symmetric, whereas the neutral gas shows pronounced small-scale structure and an asymmetric radial extent. In particular, the stellar disc extends beyond the detectable \HI\ emission on the western side of the galaxy, while on the eastern side the \HI\ and stellar distribution reach comparable projected radii.

One-sided truncations of \HI\ discs relative to the stellar distribution are often observed in systems that may be experiencing interaction with an external gaseous medium. Beyond the main stellar body, the \HI\ distribution remains structured, with localised enhancements and irregular features, whereas the stellar contours become sparse. This asymmetric gas–star configuration suggests that the low-column-density \HI\ is more strongly affected than the stellar component, consistent with expectations for environmental processes acting preferentially on the diffuse neutral gas. This may occur if the outer low-density gas is displaced by interaction with an external medium, while the stellar component remains largely unaffected due to its stronger gravitational binding. Similar offsets are also predicted in simulations of weak ram-pressure interaction (see Sect.~\ref{sec:simulations}).
\subsection{Filament analysis}\label{sec:filaments}
\begin{figure} 
\centering
   \advance\leftskip0cm
     \includegraphics[width=8.5cm]{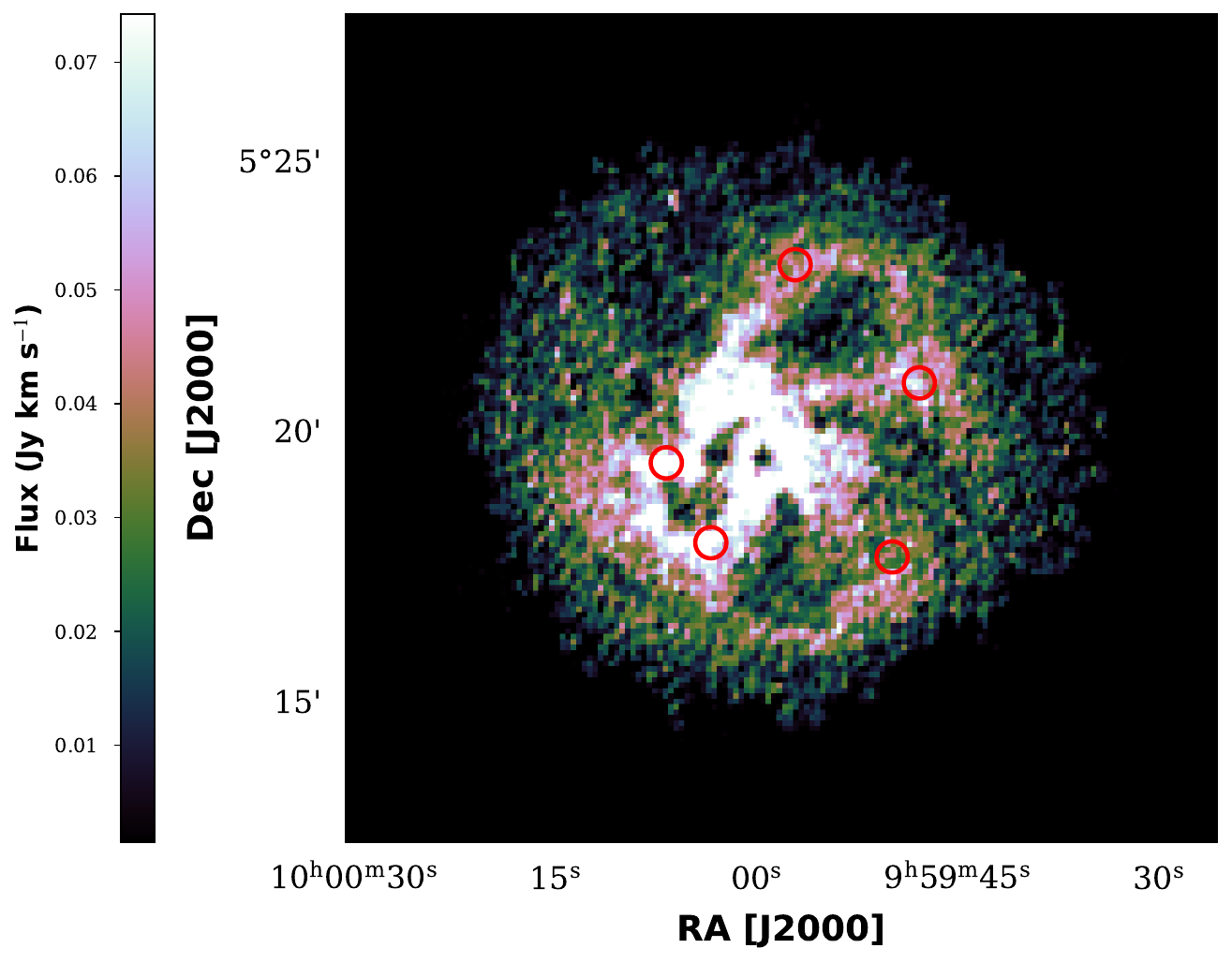}
\caption{Highest-resolution MeerKAT \HI\ integrated intensity map of Sextans~B highlighting filamentary structures. Open circle symbols mark the locations of identified \HI\ filaments, which appear as narrow, elongated extensions emerging from the outer disc. The filaments are distributed asymmetrically around the galaxy and are present only in the neutral gas, with no clear counterpart in the stellar or ionized components.}

\label{fig:filaments}
\end{figure}

Several filamentary \HI\ features are identified in the highest-resolution integrated intensity map, as indicated by star markers in Fig.~\ref{fig:filaments}. These structures appear as narrow, elongated enhancements of neutral gas extending outward from the main disc, predominantly originating near the outer edge of the high-column-density \HI\ distribution. The filaments are spatially coherent and trace localised extensions of the disc into the surrounding low-density gas. The detected filaments are distributed asymmetrically, with multiple features concentrated on one side of the galaxy, whereas the opposite side shows comparatively fewer extensions. Individual filaments exhibit lengths of several hundred parsecs and vary in morphology from nearly radial protrusions to gently curved structures. Several originate near regions of complex inner morphology, including areas adjacent to \HI\ cavities and clumpy gas concentrations. No corresponding filamentary counterparts are observed in the stellar distribution, indicating that these features are restricted to the neutral gas component.
\section{\HI\ kinematics}\label{sec:kinematics}
We derived the kinematic structure of Sextans~B using the Tilted Ring Fitting Code (TiRiFiC; \citealt{2007A&A...468..731J}). TiRiFiC performs three-dimensional modelling of the \HI\ data cube by representing the neutral gas distribution as a set of concentric tilted rings, allowing the reconstruction of the galaxy’s kinematics while accounting for projection effects.
\subsection{Modelling procedure}
In TiRiFiC, the \HI\ disc is represented by a series of concentric circular rings, each parameterised by rotation velocity, inclination, position angle, and systemic velocity. Initial estimates of these parameters were obtained using PyFAT \citep{2024ascl.soft07002K}, which performs an automated tilted-ring analysis of the data cube to derive initial estimates of the kinematic parameters from the velocity field and \HI\ surface-brightness distribution. These values were adopted as starting points for the TiRiFiC modelling.

The fitting was performed iteratively. In the initial run, the dynamical centre and systemic velocity derived from PyFAT were held fixed, while the rotation velocity was allowed to vary with radius and the inclination and position angle were kept fixed to stabilise the solution. In subsequent iterations, the inclination and position angle were gradually released to capture radial variations and possible disc warps. Radial motions were included to account for departures from purely circular rotation that may arise from disturbances in the outer disc.

To investigate kinematic asymmetries, the approaching and receding sides of the galaxy were modelled independently. Kinematic model convergence was assessed by comparing residual cubes between successive iterations and requiring that further iterations do not significantly improve the fit, with the preferred model selected based on the median absolute residual across the cube.

Uncertainties on the fitted kinematic parameters, including rotation velocity, inclination, position angle, systemic velocity, and centre position, were estimated using the \texttt{TRM-errors}\footnote{\url{https://pypi.org/project/TRM-errors/0.0.9/}} package. This tool perturbs the best-fitting TiRiFiC model via the \texttt{tirshaker} module to explore parameter space, and the resulting scatter is propagated into formal error estimates, providing a robust assessment of the model uncertainties.

\subsection{Kinematic modelling results}
\begin{figure*} 
\centering
   \advance\leftskip0cm
   \includegraphics[width=18.0cm]{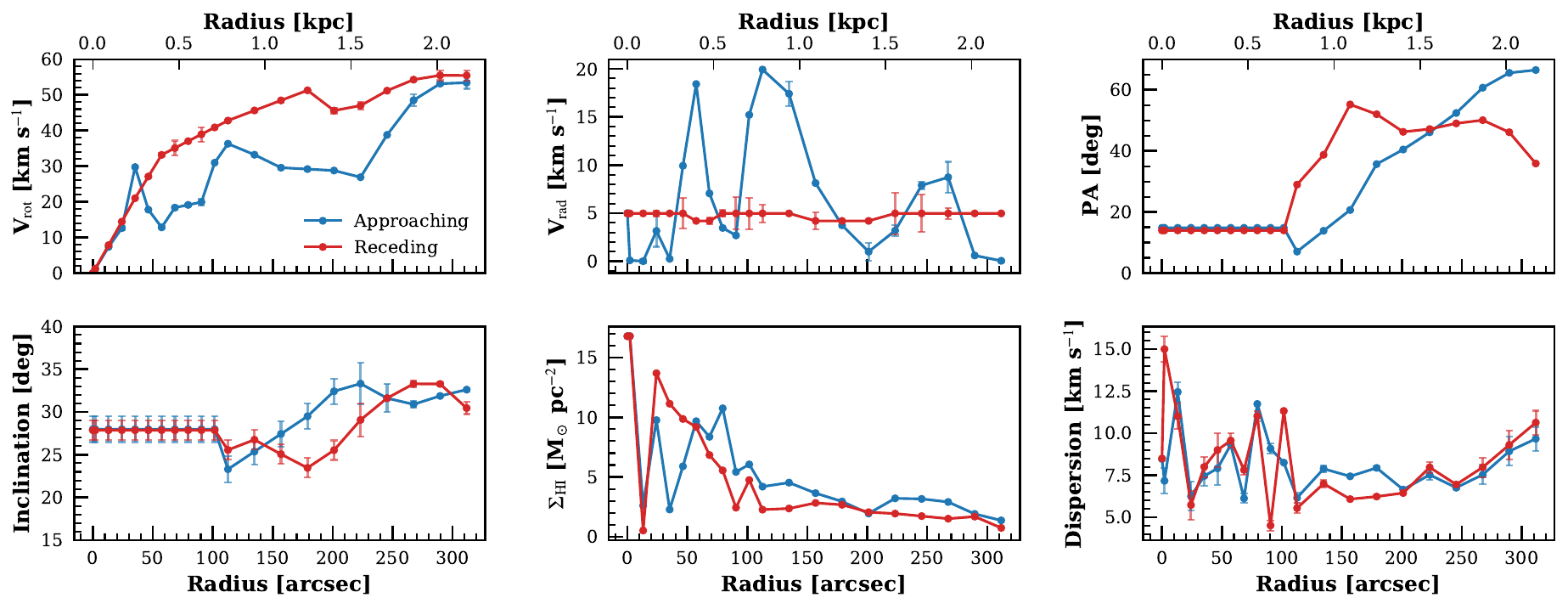}
\caption{Radial profiles derived from the TiRiFiC modelling of Sextans~B, shown separately for the approaching (blue) and receding (red) sides of the galaxy. From top to bottom: rotation velocity ($V_{\rm rot}$), radial velocity component ($V_{\rm rad}$), position angle (PA), inclination, \HI\ surface density, and velocity dispersion. Radii are shown in arcseconds (bottom axis) and kiloparsecs (top axis). Error bars represent formal uncertainties from the \texttt{TRM-errors} analysis.}
\label{fig:rc}
\end{figure*}

\begin{figure*} 
\centering
   \advance\leftskip0cm
   \includegraphics[width=14cm]{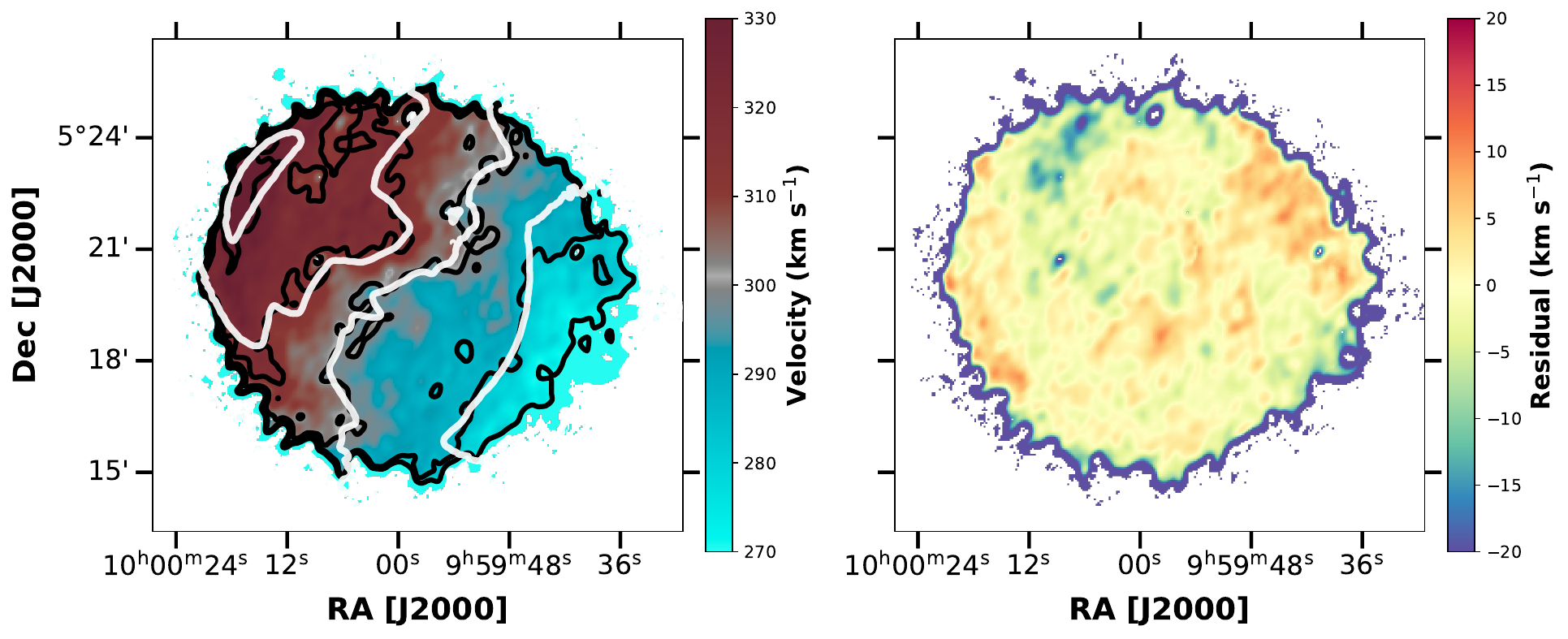}
\caption{Comparison between the observed \HI\ velocity field and the best-fit kinematic model derived from the TiRiFiC modelling.
Left: intensity-weighted velocity field (moment~1) with observed velocities shown as black contours and model velocities as white contours. Right: residual velocity field computed as the difference between the observed velocity field and the best-fit kinematic model. The model reproduces the large-scale \HI\ velocity gradient across the disc. The residuals are generally small, indicating good agreement between the data and the model.}

\label{fig:datamodelb}
\end{figure*}

\begin{figure*} 
\centering
   \advance\leftskip0cm
   \includegraphics[width=7.5cm]{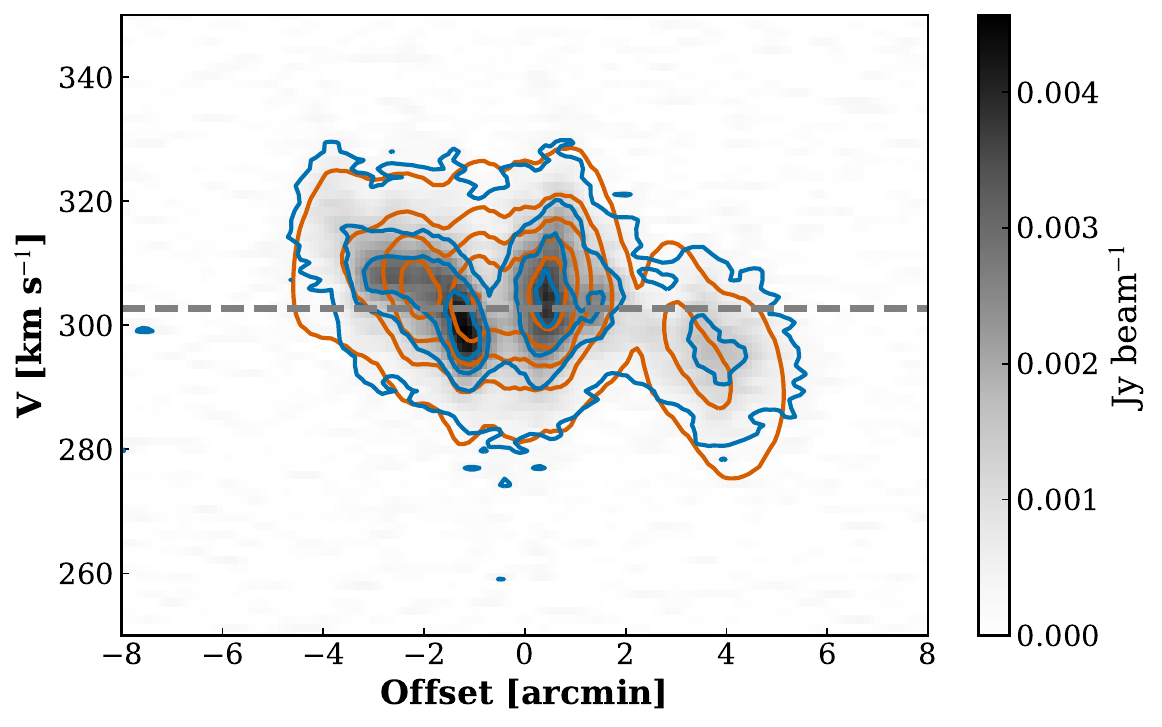}
    \includegraphics[width=7.5cm]{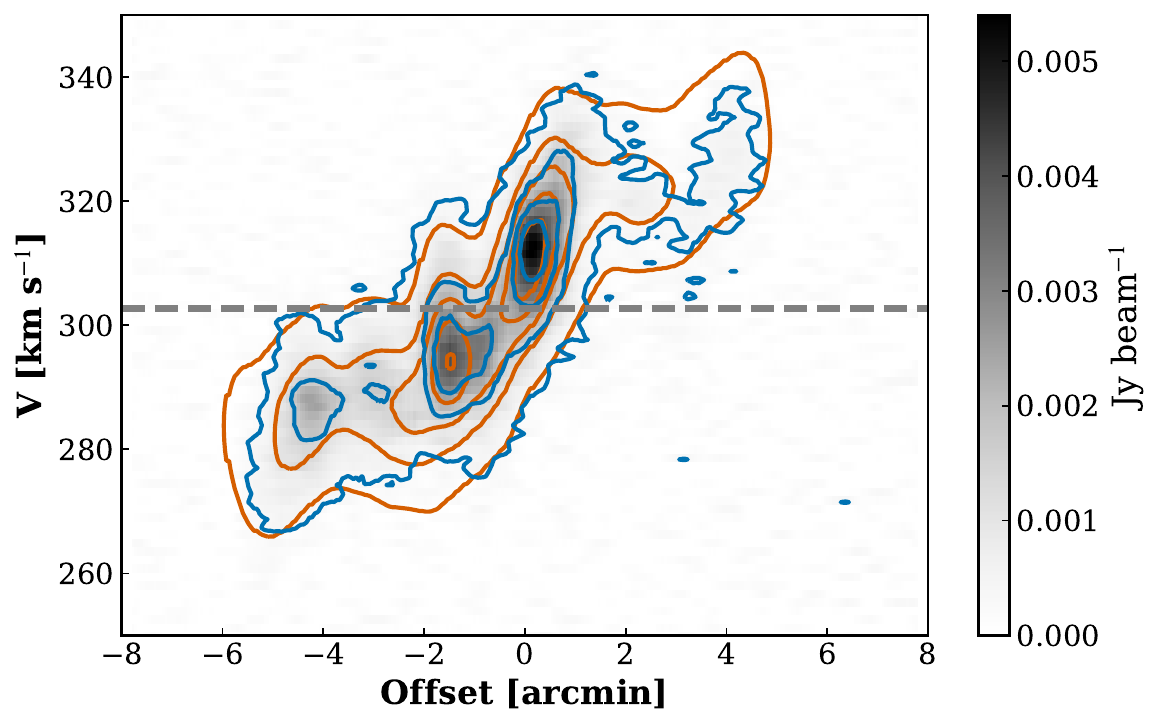}
\caption{Position--velocity diagrams of Sextans~B extracted along the kinematic major axis (left) and minor axis (right). Observed emission is shown as blue contours, with the TiRiFiC model overlaid in orange. The slices illustrate the velocity structure of the \HI\ disc and the differences between the approaching and receding sides.}
\label{fig:pvb}
\end{figure*}

The TiRiFiC kinematic modelling reproduces the global velocity gradient for Sextans~B. The results are presented in Figs.~\ref{fig:rc}--\ref{fig:pvb}. Fig.~\ref{fig:rc} presents the radial profiles derived from the TiRiFiC modelling for the approaching and receding sides of Sextans~B.

The systemic velocity derived from the kinematic modelling is $V_{\rm sys} = 302.671 \pm 0.05$ km\,s$^{-1}$. This value is consistent with the optical systemic velocity of Sextans~B ($\sim$301 km\,s$^{-1}$; e.g. \citealt{2012AJ....144....4M}).

The dynamical centre determined from the kinematic modelling is located at RA = 150.005$^\circ$ and Dec = 5.334$^\circ$. This position is slightly offset from the optical centre adopted in this work (RA = 150.000$^\circ$, Dec = 5.332$^\circ$) by approximately $\sim17$ arcsec. Such small offsets between optical and kinematic centres are not uncommon in dwarf irregular galaxies and may reflect asymmetries in the neutral gas distribution. The measured offset ($\sim17$ arcsec) is larger than the beam size of the data cube ($\sim10$ arcsec), and is therefore spatially resolved.

The rotation curves rise rapidly within the inner $\sim$0.5~kpc and then gradually increase, reaching velocities of $\sim$50--55~km\,s$^{-1}$ at the outermost radii. The receding side generally exhibits higher rotation velocities than the approaching side at intermediate radii, indicating a clear kinematic asymmetry between the two sides of the outer disc. 

The radial velocity component remains small over much of the disc but exhibits localised deviations on both sides, with amplitudes reaching up to $\sim$20~km\,s$^{-1}$ at several radii. The position angle varies with radius, particularly beyond $\sim$100~arcsec, where the approaching and receding sides follow distinct radial trends. The inclination remains close to $\sim$30$^\circ$ but shows moderate radial variations. The \HI\ surface density declines steadily with radius on both sides, with noticeable differences in amplitude and radial structure between the approaching and receding profiles.

Fig.~\ref{fig:datamodelb} compares the observed and modelled moment~1 maps of Sextans~B. The TiRiFiC kinematic model reproduces the large-scale velocity gradient, providing a good match to the global kinematic structure of the galaxy. Differences between the data and the kinematic model are primarily confined to localised regions of the inner disc, coincident with areas of complex \HI\ morphology, with residual velocities typically remaining below $\sim$10~km\,s$^{-1}$.

Position--velocity slices extracted along the kinematic major and minor axes (Fig.~\ref{fig:pvb}) further demonstrate that the kinematic model successfully recovers the overall rotation pattern while also reproducing the asymmetry between the approaching and receding sides. Together, these results indicate that while the inner disc remains largely regular, the outer \HI\ disc shows clear kinematic asymmetry between the approaching and receding sides.

\subsection{Asymmetric drift correction}
\begin{figure}
\centering
\includegraphics[width=8.5cm]{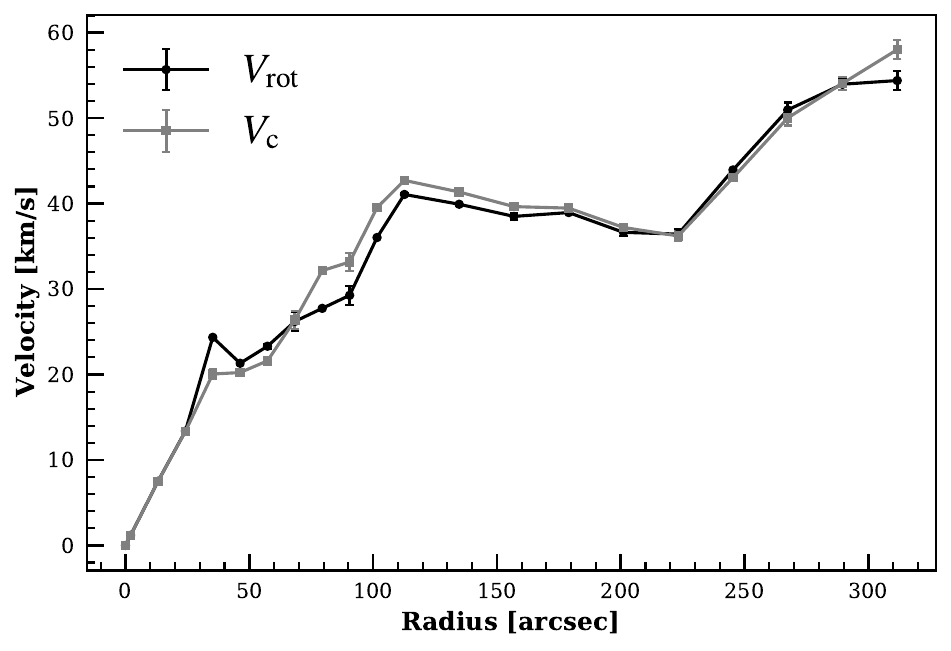}
\caption{Averaged rotation curve of Sextans~B before and after asymmetric drift correction. Black points show the rotation velocities derived from the TiRiFiC modelling, while grey points indicate the corresponding circular velocities after correcting for pressure support.}
\label{fig:asy}
\end{figure}

We applied an asymmetric drift correction to account for dynamical support from random gas motions. Following \citet{1996AJ....111.1551M}, the circular velocity $V_{\rm c}$ is related to the observed rotation velocity $V_{\rm rot}$ by
\begin{equation}
    V_{\rm c}^2(R) = V_{\rm rot}^2(R) + \sigma_{D}^2(R),
\end{equation}
where the drift term $\sigma_{D}^2$ is given by
\begin{equation}
    \sigma_{D}^2(R) = -\,R\,\sigma^2(R)
    \left[ \frac{\partial \ln \Sigma_{g}(R)}{\partial R}
         + 2\,\frac{\partial \ln \sigma(R)}{\partial R}
         - \frac{\partial \ln h_{z}(R)}{\partial R} \right].
\end{equation}
Here $V_{\rm c}$ is the circular velocity, $V_{\rm rot}$ is the observed rotation velocity, $\Sigma_{g}$ is the gas surface density (including helium), $R$ is the galactocentric radius, $\sigma$ is the velocity dispersion, and $h_{z}$ is the vertical scale height of the gas disc. The velocity dispersion profile $\sigma(R)$ was taken from the best-fitting TiRiFiC model. In computing the correction, we assume that the radial variation of the scale height is negligible, i.e. $\partial \ln h_{z} / \partial R = 0$. This assumption is commonly adopted in asymmetric-drift corrections (e.g. \cite{1996AJ....111.1551M}), since the correction depends only on the radial derivative of $\ln h_z$. Studies of dwarf galaxies show that the H\,I scale height varies only gradually with radius (e.g. \cite{2011MNRAS.415..687B}), while the surface density typically changes more strongly, so the correction is dominated by the $\Sigma_g$ and $\sigma$ terms. The resulting asymmetric drift–corrected rotation curve for Sextans~B after averaging both sides is shown in Fig.~\ref{fig:asy}.
\section{Comparison with ram-pressure simulations}\label{sec:simulations}
\subsection{The IGM wind tunnel}
To interpret the \HI\ ~ morphology and kinematics of Sextans~B, we compare our observations to a hydrodynamical ram-pressure simulation originally developed for gas-rich dwarf irregular galaxies moving through a low-density intergalactic medium by \citet{2022A&A...660L..11Y} using the hydrodynamics code GIZMO \citep{GIZMO}. These simulations model a rotating dwarf galaxy embedded in a dark-matter halo and are subjected to an idealized ambient medium representing the motion of the galaxy through the diffuse gas of the Local Group. A spatial cube with a side of length 500 kpc is initiated with a uniform density IGM with a mean temperature of 1 million Kelvin. GIZMO allows for a "periodical" configuration in which the cube is spatially looped, where a particle trying to cross one face of the cube, appears on the opposite side. This serves as an extended "wind tunnel" for the galaxy to move through, making it ideal for studying the impact of the IGM on isolated galaxies. Within this framework, we executed a simulation tailored to reproduce and interpret some features seen in Sextans B. For the galaxy itself, we use an exponential profile for the stellar and gaseous discs with a scale length of 1~kpc and a scale height of 0.3~kpc. The discs are embedded in a cored spherical dark matter halo with a scale length of 1.2~kpc. The total stellar mass for the initial condition is $9.5 \times 10^{6}\,M_\odot$, $1.44 \times 10^{8}\,M_\odot$ for the gaseous disc (including helium), and the total mass of the halo is $1.5 \times 10^{9}\,M_\odot$. We adopted a particle mass of $1375 M_\odot$ for the baryonic particles and $5512.4M_\odot$ for the dark matter particles. The disks and the halo are generated by using ZENO \citep{zeno}.

Incorporating realistic star formation and feedback prescriptions is essential for studying N-body simulations of star forming gas-rich irregulars. In our simulation, the star formation rate is tuned using the volumetric star formation laws for dwarf galaxies from \citet{Volumetric_SFR}. Prescriptions for feedback follow those described by \citet{2006MNRAS.373.1013C}, which were successfully used by \citet{Wang2024} in GIZMO. With realistic inputs for both, our simulation accurately converts gas particles to stellar particles with time, and injects energy mimicking star formation and supernovae.

\subsection{Constraints from observations}
During our modelling, we face a degeneracy between the input velocity through the IGM and the density of the latter. We choose to constrain these parameters by using previously reported limits for group environments in other studies. \citet{RPS_pegdig} and \citet{2022A&A...660L..11Y} constrain the Local Group IGM density to $10^{-6}$--$10^{-5}\,\mathrm{cm^{-3}}$ by studying interaction of Pegasus dwarf irregular galaxy (DDO 216) and WLM with the IGM. Similarly, \citet{2002AJ....123.1316B} find similar values of $4 \times 10^{-6}$~atoms~cm$^{-3}$ for the M81 group. Taking these observational constraints into account, we adopt an IGM density of $4 \times 10^{-6}$~atoms~cm$^{-3}$ in our simulation. Additionally, we also choose to run a separate simulation where the same Initial condition of the galaxy evolves in complete isolation, with no interaction with the IGM as a benchmark to compare morphological differences in the gas disk.

The motion and velocity of the NGC 3109 association, of which Sextans~B is reported to be a part of has been of great interest in studies focussing on global motions of dwarfs in the Local Group \citep{Perseus_I_NGC_3109,NGC_3109_backsplash}. Sextans~B has a radial velocity of 171 ~km\,s$^{-1}$ with respect to the MW and a systemic velocity of 300 ~km\,s$^{-1}$ \citep{2012AJ....144....4M}. Though \citet{2022A&A...657A..54B} obtained proper motions for Sextans~B, its distance is so large (1.36 Mpc) that it is quite dominated by error bars rendering it difficult to estimate an accurate 3D space velocity. They highlight this by mentioning an error of 1000 ~km\,s$^{-1}$ on the transverse velocity. However, considering the space velocity of other isolated irregulars such as WLM is $300 \pm 150$ ~km\,s$^{-1}$ \citep{2022A&A...660L..11Y} and IC 10 is $215 \pm 42$ ~km\,s$^{-1}$ \citep{IC10_proper_motion} and with a constrained radial velocity of Sextans~B, it is reasonable to compare our Sextans~B observations to a  hydrodynamical model of a gas-rich irregular moving with a space velocity of a few 100s of ~km\,s$^{-1}$, therefore our model moves at a velocity of 300 ~km\,s$^{-1}$ through the IGM. Taking into account the large radial velocity of motion away from the MW, and Sextans~B's inclination being close to 30$^\circ$, The initial condition of the galaxy is placed with the disk tilted 30$^\circ$ into the
medium with a leading edge in the direction of motion through the IGM. Thus, we have a model of a gas rich irregular moving away from the observer through the IGM with the observer looking at the face of the disk as it recedes. These observationally motivated constraints let us compare this illustrative hydrodynamical model to our Sextans~B observations. Future limits on proper motions with the next GAIA data release will significantly help later studies, though the bulk motion of the local IGM, if any, so far remains fully unconstrained. 
\subsection{Evolution of the gas disk in the IGM}
As the simulation begins the galaxy traverses through the medium through the 500 kpc wide cube. The simulation runs for 10 gigayears in total. Within the first gigayear, the galaxy, particularly the gas disk, rapidly evolves away from the initial conditions due to star formation and feedback. As the galaxy evolves further, we see that even weak ram pressure, characteristic of group environments, can substantially perturb the outer H i disc while leaving the inner regions largely unaffected. Low-column-density gas is displaced from the outer edges of the disk and the velocity field develops systematic differences between the approaching and receding sides beyond $\simeq$1--2~kpc. In this regime, the outer disc becomes progressively out of dynamical equilibrium, whereas the stellar component and the dense inner gas remain comparatively stable.  

\begin{figure*}[h!] 
\centering
   \advance\leftskip0cm
     \includegraphics[width=19cm]{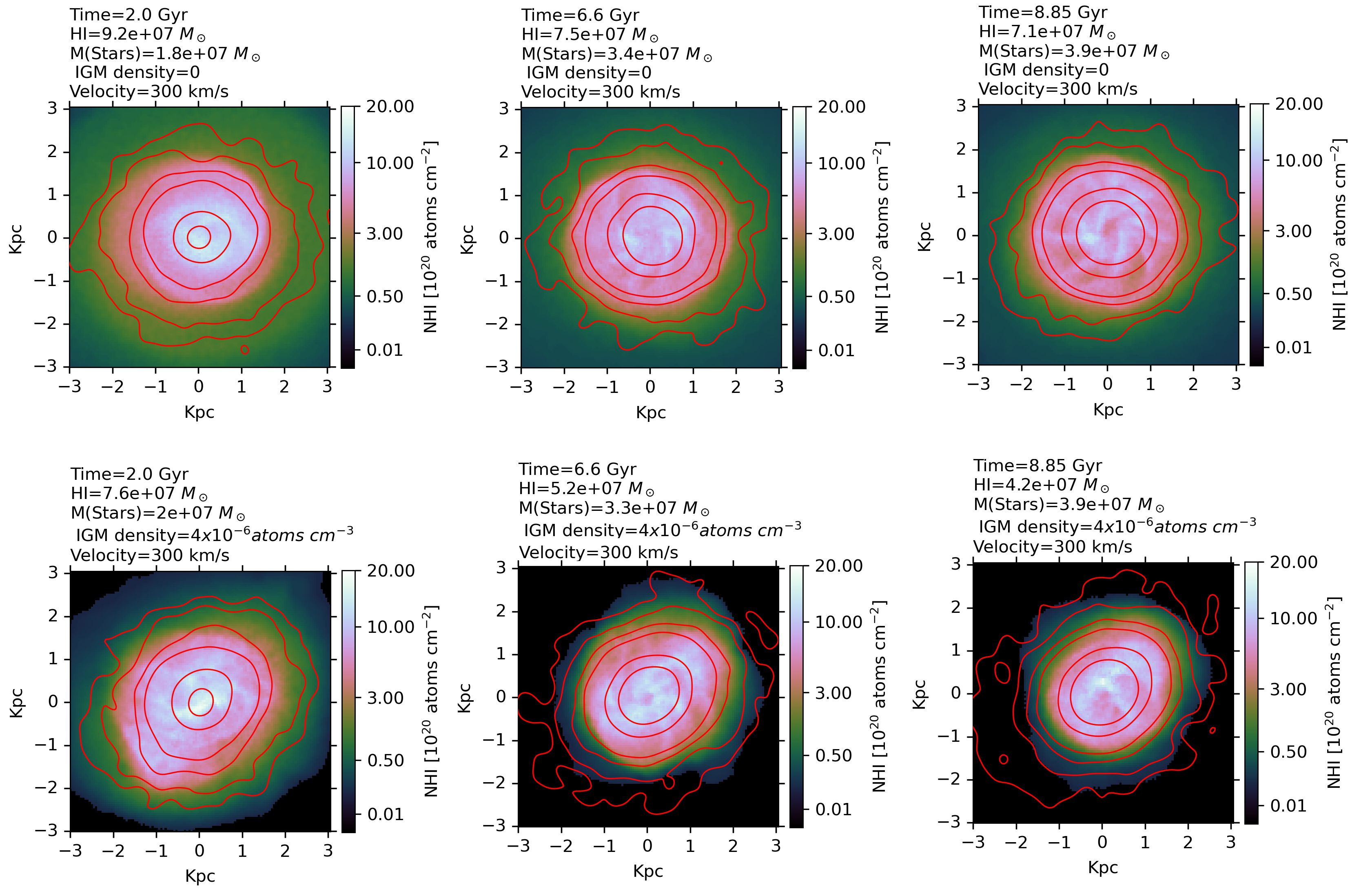}
\caption{Evolution of the simulated galaxy. In the top row, we present the evolution of the simulated galaxy described in section \ref{sec:simulations} at 3 different epochs as the galaxy evolves in complete isolation without interacting with the IGM, no gas is stripped and the gas disk symmetrically envelopes the stellar disc at all epochs. In the bottom row we present three snapshots at the same epochs as the top row showing the evolving morphology of the galaxy as it interacts with the IGM. It loses gas and tends towards a gaseous and stellar distribution similar to the observations. Column density maps are overlaid with stellar mass surface density contours of values 0.1,0.2,0.4,0.8,3.1,6.3 $M_\odot/pc^{-2}$ from the outside in. Bottom left: 2 Gyr into the IGM, the gaseous disk extends beyond the stellar disk. Bottom center: 4.6 Gyr later, we have an asymmetry in the gas and stellar distribution. Bottom right: After losing gas to ram pressure for 8.85 Gyr we arrive at an asymmetrical distribution of \HI\ and the stellar disk, they are clearly offset in a similar fashion to the observations. The internal filamentary structure seen in Sextans~B is seen at all epochs, though the system under ram pressure seems to have evolved larger, asymmetric ones.}

\label{fig:disk_evo}
\end{figure*}

We place our simulation within an observation-like frame (PA, inc) so that we can compare our \HI\ ~ observations to the simulations robustly. In Fig. {\ref{fig:disk_evo}}, on the top we see an instance of our galaxy initial condition evolving without interacting with the IGM, with no gas loss. The bottom row in the same figure illustrates what happens to the disk as it evolves through the IGM and develops an asymmetric gas distribution with time. A feature to note between the instance in isolation and one evolving in the IGM, is the distinct oval-like shape of the inner disk, while the one in isolation remains nearly circular. The one interacting with the IGM also distinctly shows a few, broad filaments, similar to those in the observations, compared to the numerous small ones seen in the one is isolation at the 8.85 Gyr epoch. 

\begin{figure*}[h!] 
\centering
   \advance\leftskip0cm
     \includegraphics[width=19cm]{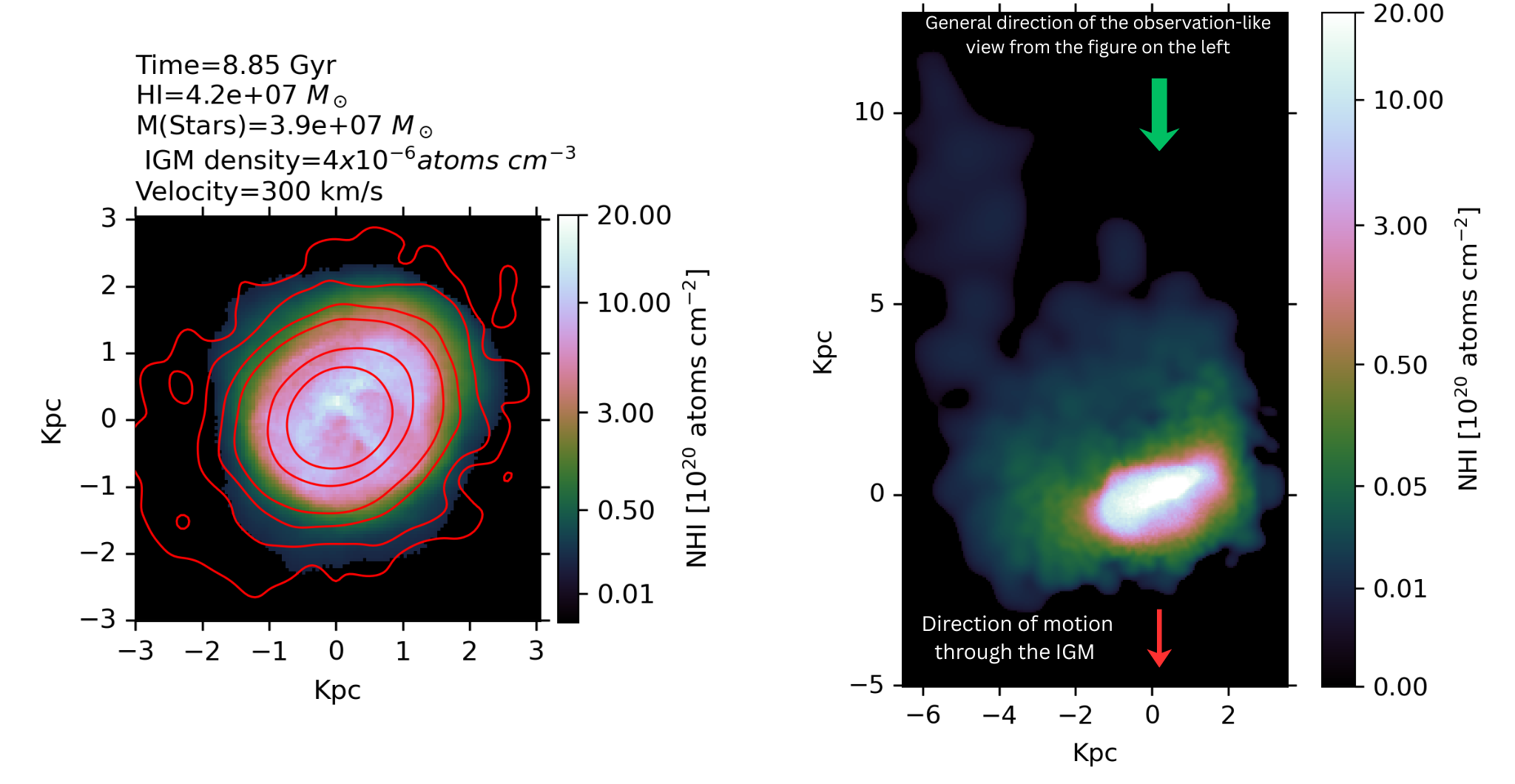}
\caption{Simulated dwarf galaxy evolving through the IGM after 8.85 Gyr. Left: A column density map of a simulated dwarf galaxy like Sextans~B evolving through the IGM after 8.85 Gyr, reaching a total stellar mass and HI mass very similar to the observed values. The contours represent stellar mass surface density with values of 0.4, 0.6, 0.8, 1.6, 3.0, 9.4, 12.6 $M_\odot/pc^{-2}$ from the outside in. We can see that similar to the observations presented in Fig. \ref{fig:high} panel(b), the stellar contours extend beyond the HI envelope towards the southern edge. The units used for the stellar contours in Fig. \ref{fig:high} panel(b) differ from those presented here, but serve an illustrative purpose to demonstrate an offset in the stellar and gaseous distributions. Right: Edge on HI column density map of same galaxy as on the left at the same epoch, evolving through a sparse IGM density of $4 \times 10^{-6}$~atoms~cm$^{-3}$ and steadily losing gas as described in section \ref{sec:simulations}. This snapshot is the same as the one presented in the bottom right panel of Fig. {\ref{fig:disk_evo}}.}
\label{fig:front-edge}
\end{figure*}

\begin{figure*} 
\centering
   \advance\leftskip0cm
   \includegraphics[width=18.0cm]{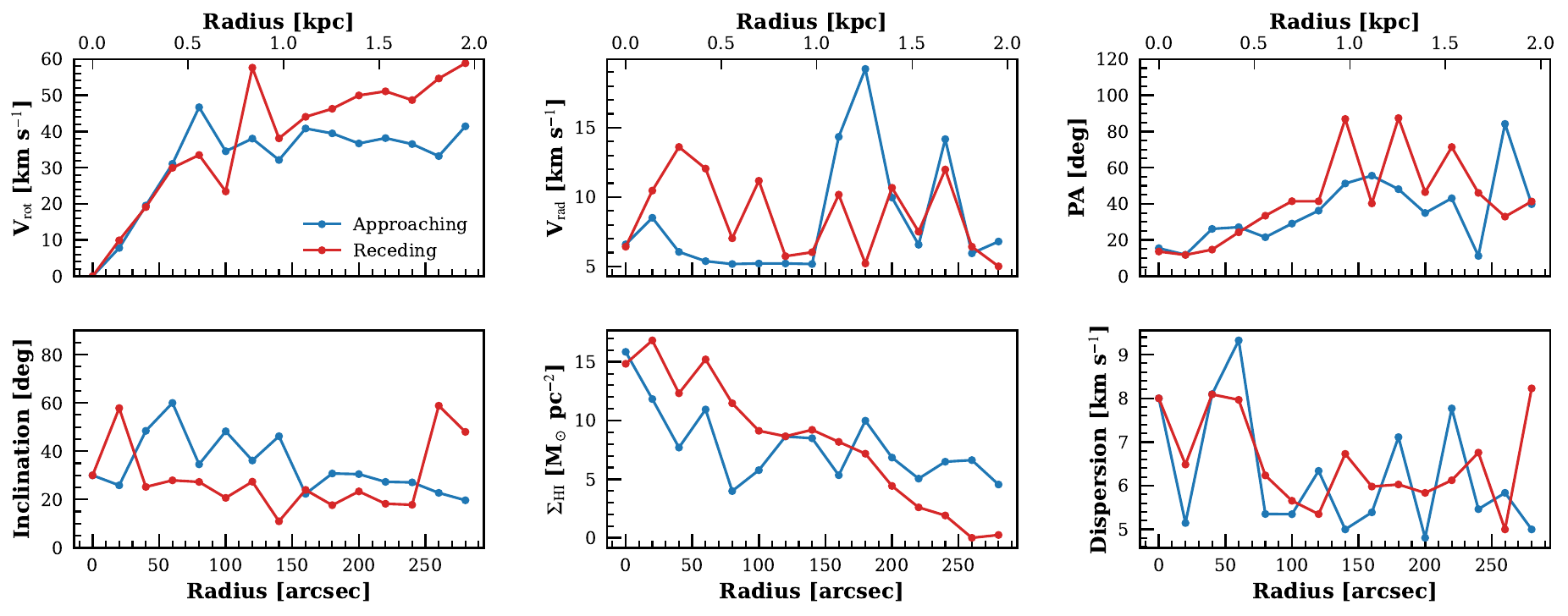}
\caption{Radial profiles derived from the TiRiFiC modelling of the \HI ~ disk of the hydrodynamical simulation presented in Fig.\ref{fig:disk_evo} at  8.85 Gyr, shown separately for the approaching (blue) and receding (red) sides of the galaxy. From top to bottom: rotation velocity ($V_{\rm rot}$), radial velocity component ($V_{\rm rad}$), position angle (PA), inclination, \HI\ surface density, and velocity dispersion. Radii are shown in arcseconds (bottom axis) and kiloparsecs (top axis).}
\label{fig:simrc}
\end{figure*}

The image presented on the left in Fig. {\ref{fig:front-edge}} is a snapshot of the simulation in the observational frame, which has evolved 8.85 Gyr in the IGM, the figure on the right in Fig. \ref{fig:front-edge} is the same simulation seen from an edge-on frame, to view how the galaxy is losing its gas. The tilt of the galaxy with respect to the IGM can be seen here. 
TiRiFiC radial profiles for this epoch are presented in in Fig. {\ref{fig:simrc}}.


\section{Discussion}\label{sec:discussion}
The deep MeerKAT observations presented in this work reveal several observational signatures suggesting that the outer neutral gas disc of Sextans~B is dynamically perturbed. These include asymmetries in the low--column-density \HI\ distribution, spatial differences between the stellar and gaseous components, and systematic discrepancies between the approaching and receding sides of the rotation curve derived from the kinematic modelling. Taken together, these morphological and kinematic properties indicate that the outer gas disc departs from a simple symmetric configuration.

A similar combination of disturbed \HI\ morphology and asymmetric outer-disc kinematics has recently been reported for the isolated Local Group dwarf galaxy WLM, where interaction with a diffuse ambient medium has been proposed to explain both the extended \HI\ structures and the divergence between the approaching and receding rotation curves  \citep{2022A&A...660L..11Y,Kolhe2026}. Within this emerging framework, Sextans~B provides an additional system in which to investigate whether weak ram-pressure interaction with a low-density intergalactic medium can influence both the structure and kinematics of dwarf galaxies in the outskirts of the Local Group. 
\subsection{Observational signatures of environmental interaction}
The kinematic modelling presented in this work provides further evidence that the outer gas disc of Sextans~B is dynamically perturbed. The rotation curves derived for the approaching and receding sides of the galaxy diverge at larger radii, while the inner regions remain comparatively regular. This behaviour suggests that the outer velocity field deviates from simple axisymmetric rotation.

While stellar feedback is known to produce cavities, shells, and small-scale filamentary structures in the neutral gas of dwarf galaxies \citep{1992AJ....103.1841P,1999AJ....118..273W}, it typically operates on local scales and is not generally expected to generate large-scale one-sided distortions of the outer \HI\ envelope or systematic kinematic differences between the two sides of the disc. The combination of asymmetric outer \HI\ morphology, gas–star offsets, and diverging rotation curves therefore suggests that internal feedback alone may not fully account for the observed features, and that an additional external perturbation may be affecting the outer gas disc. Although asymmetric rotation curves can also arise from internal processes such as disc warps, lopsided mass distributions, or non-circular gas motions, the coincidence between the kinematic asymmetry and the morphological distortions observed in the outer \HI\ disc suggests that both may originate from a common perturbation acting preferentially on the low-density outer gas.

A similar combination of disturbed \HI\ morphology and asymmetric outer-disc kinematics has recently been identified in the Local Group dwarf galaxy WLM \citep{2007AJ....133.2242K,2022A&A...660L..11Y,Kolhe2026}. In that system, the divergence between the approaching and receding rotation curves at large radii has been interpreted as evidence for ram-pressure effects acting on the outer gas disc. Taken together, the combination of asymmetric \HI\ morphology, gas--star offsets, filamentary structures, and diverging rotation curves forms a consistent set of observational signatures that may indicate interaction between the outer gas disc of Sextans~B and a diffuse ambient medium.

\subsection{Interpretation from ram-pressure simulations}
We have performed hydrodynamical simulations of a Sextans~B--like dwarf galaxy evolving in a low-density intergalactic medium representative of the Local Group environment, following the prescriptions described in Section~\ref{sec:simulations}. These simulations show that the combined effects of ram pressure and stellar feedback can naturally generate asymmetric gas distributions and filamentary structures in dwarf galaxies \citep{2013ApJ...763L..41B}.

In the early stages of the simulated evolution, star formation and stellar feedback create cavities and filamentary gas structures within the disc. As the galaxy continues to interact with a diffuse ambient medium, ram pressure gradually removes gas from the outer regions of the disc. This interaction leads to an asymmetric outer gas distribution. A demonstration of this process can be seen in Fig.\ref{fig:disk_evo}.

The efficiency of this process depends on the balance between the external pressure and the gravitational restoring force of the galaxy, commonly expressed through the classical condition proposed by \citet{1972ApJ...176....1G}. The ram pressure exerted by the ambient medium can be written as
\begin{equation}
P_{\rm ram} = \rho_{\rm IGM} v^2 ,
\end{equation}
where $\rho_{\rm IGM}$ is the density of the surrounding medium and $v$ is the relative velocity between the galaxy and the ambient gas. Gas stripping is expected to occur when the ram pressure exceeds the gravitational restoring force per unit area of the disc, which can be approximated by
\begin{equation}
P_{\rm grav} \approx 2 \pi G \Sigma_* \Sigma_g ,
\end{equation}

where $\Sigma_*$ and $\Sigma_g$ are the stellar and gas surface densities, respectively \citep{1972ApJ...176....1G}. For an ambient density of $n \sim4\times10^{-6}$ atoms cm$^{-3}$ and a velocity of $\sim300$ km s$^{-1}$, the corresponding ram pressure is of order $P_{\rm ram} \sim10^{-15}\,\mathrm{dyn\,cm^{-2}}$. Using the \HI\ surface density derived from our TiRiFiC modelling of the observations and adopting outer-disc stellar surface densities from \citet{2015AJ....149..180O}, we estimate $P_{\rm grav} \sim \mathrm{few} \times 10^{-15}\,\mathrm{dyn\,cm^{-2}}$, indicating that the ram pressure is sufficient to perturb the outer low-density gas without fully stripping the disc.

The initial condition of the simulation had a total stellar mass of $9.6 \times 10^{7} M_\odot$ and a gas mass of $1.44 \times 10^{8} M_\odot$. Assuming a neutral hydrogen fraction of 0.76 therefore, the total HI mass was $1.1\times 10^{8} M_\odot$. At the 8.85 Gyr epoch, the total stellar mass is  $3.9\times 10^{7} M_\odot$ and the total \HI\ ~  mass in the disk is $4.2\times 10^{7} M_\odot$ , thus only $2.94\times 10^{7} M_\odot$ of the HI mass was converted to new stars, $3.86\times 10^{7} M_\odot$ was lost to the combined effect of ram pressure and feedback. Because dwarf galaxies possess shallow gravitational potentials and extended low-density \HI\ discs, even relatively weak external pressures can influence their outer gas distributions \citep{2002AJ....123.1316B}

Radial profiles of the model at 8.85 Gyr derived using TiRiFiC presented in Fig. \ref{fig:simrc}. While the simulations are idealised and not specifically tailored to reproduce the exact parameters of Sextans~B, the radial profiles indicate that our model is morphologically and kinematically similar enough to Sextans-B  to illustrate the expected signatures of ram-pressure interaction which can be seen in Fig.\ref{fig:disk_evo}.

In the simulations presented in Section~\ref{sec:simulations}, a dwarf galaxy moving through an intergalactic medium with density $\sim4\times10^{-6}$ atoms cm$^{-3}$ and velocity $\sim300$ km s$^{-1}$ develops extended asymmetric gas structures and outer-disc kinematic perturbations that qualitatively resemble the features observed in Sextans~B.

\subsection{Implications for dwarf galaxies in low-density environments}
The case of Sextans~B supports the emerging view that environmental processes may influence dwarf galaxies even in relatively low-density regions of the Local Group. Observations of the isolated dwarf galaxy WLM reveal extended \HI\ clouds trailing behind the galaxy and kinematic asymmetries in its rotation curve, both interpreted as signatures of interaction with a diffuse intergalactic medium \citep{2022A&A...660L..11Y,Kolhe2026}. The simulation shown in Fig.~9 corresponds to a dwarf galaxy evolved in an intergalactic medium with density $\sim4\times10^{-6}$ atoms cm$^{-3}$ and a relative velocity of $\sim300$ km s$^{-1}$ (see Section~\ref{sec:simulations}).

Because the ram pressure expected in such low-density environments is relatively weak, its influence is likely to accumulate gradually over several gigayears rather than producing rapid gas stripping as observed in dense cluster environments. Over these long timescales, even mild hydrodynamical interactions can progressively reshape the outer \HI\ discs of dwarf galaxies and alter their kinematics while leaving the stellar component largely unaffected.

Sextans~B therefore provides an additional example supporting the possibility that weak environmental effects may contribute to shaping the gas distribution and dynamics of dwarf galaxies located in the outskirts of the Local Group.

\section{Conclusions}\label{sec:conclusion}
We have presented deep MeerKAT \HI\ observations of the dwarf irregular galaxy Sextans~B, enabling a detailed investigation of the morphology and kinematics of its neutral gas disc down to low column densities. By combining the analysis of the \HI\ distribution with three-dimensional kinematic modelling and comparisons with hydrodynamical simulations, we explored whether the observed properties of Sextans~B are consistent with weak environmental effects in the outskirts of the Local Group. Our main results can be summarised as follows:
\begin{enumerate}
\item The \HI\ distribution of Sextans~B becomes increasingly asymmetric at low column densities, revealing filamentary structures and diffuse extensions in the outer disc. These features depart from a simple symmetric configuration and suggest that the outer gas disc is dynamically disturbed.
\item A comparison between the stellar and gaseous components shows that while the stellar distribution remains relatively symmetric, the \HI\ disc exhibits clear asymmetries and spatial offsets. Such gas–star mismatches are consistent with the behaviour expected when the gaseous component is more strongly affected by external hydrodynamical forces.
\item Three-dimensional kinematic modelling of the \HI\ cube reveals systematic differences between the approaching and receding sides of the galaxy. While the inner disc remains largely regular, the rotation curves diverge in the outer regions, indicating that the outer kinematics departs from simple axisymmetric rotation.
\item The combination of asymmetric \HI\ morphology and outer-disc kinematic perturbations resembles the signatures recently reported for the isolated Local Group dwarf galaxy WLM, where environmental interaction has been proposed to explain both the disturbed gas morphology and the asymmetric rotation curve.
\item Hydrodynamical simulations demonstrate that ram-pressure effects, combined with stellar feedback, can reproduce several of the morphological features observed in Sextans~B. Over long timescales such processes can gradually reshape the outer gas distribution of dwarf galaxies while leaving the stellar component largely intact.
\end{enumerate}
These results suggest that Sextans~B may represent an additional system in which the outer gas disc of a dwarf galaxy may be influenced by interaction with a diffuse ambient medium. If confirmed in other systems, such signatures would indicate that even the extremely low-density intergalactic medium may contribute to shaping the gas distribution and kinematics of dwarf galaxies located in the outskirts of the Local Group. Future deep \HI\ observations with facilities such as MeerKAT and upcoming SKA surveys will provide the sensitivity required to probe the low-column-density gas in nearby dwarf galaxies and will be crucial for determining how common such environmental signatures are in the Local Group.
\begin{acknowledgements}
The MeerKAT telescope is operated by the South African Radio Astronomy Observatory, which is a facility of the National Research Foundation, an agency of the Department of Science and Innovation. BN, LVM, RI, MK, and AS acknowledge financial support from the grant CEX2021-001131-S funded by MICIU/AEI/10.13039/501100011033, and from the grants PID2021-123930OB-C21 and PID2024-155817OB-I00 funded by MICIU/AEI/10.13039/501100011033 and by ERDF/EU. BN also acknowledges financial support from the grant PTA2023-023268-I funded by MICIU/AEI/10.13039/501100011033 and by ESF+ and the grant DGP\_POST\_2024\_01021, funded by the Junta de Andalucía/CUII and the ESF+. MK also acknowledges that this project has received funding through the SAFE -- ``Supporting At-Risk Researchers with Fellowships in Europe'' project, funded by the European Union under Grant Agreement No.\ 101148426. The views and opinions expressed are those of the author(s) and do not necessarily reflect those of the European Union. HC acknowledges the financial support from Zhejiang Provincial Natural Science Foundation of China (Grant No. LY24A030001) and the Leading Innovation and Entrepreneurship Team of Zhejiang Province of China (Grant No. 2023R01008). Simulations in this work were performed at the High-performence calculation (HPC) resources MesoPSL financed by the project Equip@Meso (reference ANR-10-EQPX- 29-01) of the program “Investissements d’Avenir” supervised by the ‘Agence Nationale de la Recherche’.
\end{acknowledgements}
\bibliographystyle{aa}

\end{document}